\documentclass[letterpaper,12pt]{amsart}
\usepackage{latexsym,array,delarray,amsthm,amssymb,epsfig,tabu}
\usepackage{caption}
\usepackage{diagbox}
\usepackage{caption}
\usepackage{subcaption}
\usepackage[]{algorithm2e}
\usepackage{hyperref}

\theoremstyle{plain}
\newtheorem{thm}{Theorem}[section]

\newtheorem*{thm*}{Theorem}
\newtheorem*{lemma*}{Lemma}
\newtheorem*{prop*}{Proposition}
\newtheorem*{cor*}{Corollary}
\newtheorem*{conj*}{Conjecture}

\theoremstyle{definition}
\newtheorem{defn}[thm]{Definition}
\newtheorem{ex}[thm]{Example}

\theoremstyle{remark}

\newcommand{\ind}{\mbox{$\perp \kern-5.5pt \perp$}}

\newcommand{\im}{\text{im}}
\newcommand{\R}{{\tt{R}}}

\makeatletter
\providecommand\@dotsep{5}

\begin{document}

\title{Statistical Learning with Phylogenetic Network Invariants}

\author{Travis Barton}
\address{Pivotal Life Sciences }
\email{travis.barton@pivotallifesciences.com  }

\author{Elizabeth Gross}
\address{University of Hawai`i at M\={a}noa}
\email{egross@hawaii.edu}

\author{Colby Long}
\address{The College of Wooster}
\email{clong@wooster.edu}

\author{Joseph Rusinko}
\address{Hobart and William Smith Colleges}
\email{rusinko@hws.edu}


\begin{abstract}
Phylogenetic networks provide a means of describing the evolutionary
history of sets of species believed to have undergone hybridization 
or gene flow during their evolution. 
The mutation process for a set of such species can be modeled
as a Markov process on a phylogenetic network.
Previous work has shown that a site-pattern probability
distributions from a Jukes-Cantor phylogenetic network model must
satisfy certain algebraic invariants. As a corollary,
aspects of the phylogenetic network are theoretically 
identifiable from site-pattern frequencies. In practice,
because of the probabilistic nature of sequence evolution,
the \emph{phylogenetic network invariants} will rarely be
satisfied, even for data generated under the model. 
Thus, using network invariants
for inferring phylogenetic networks requires
some means of interpreting the residuals, or deviations from zero, 
when observed site-pattern frequencies are substituted into the invariants.
In this work, we propose a method of
utilizing invariant residuals and support vector machines to infer 4-leaf level-one phylogenetic networks, from which larger networks can be reconstructed.
Given data for a set of species, the support vector machine is first trained on model data to learn the patterns of residuals corresponding to
different network structures to classify the network that produced
the data. We demonstrate the performance of our method on simulated data from the specified model and primate data.  
\end{abstract}
 
\maketitle

Phylogenetic networks are directed acyclic graphs that aim to describe the evolutionary relationships among a set of taxa. Less restrictive than their tree counterparts, phylogenetic networks have the flexibility to model gene flow and reticulation events such as hybridization and horizontal gene transfer. Due to this flexibility, phylogenetic networks are becoming increasingly common in phylogenetic analysis, and new tools are needed for their inference. In this work, we approach the inference problem from an algebro-geometric framework, combining tools from computational algebraic geometry and statistical learning.
 
Since phylogenetic network inference from genetic data is still relatively new, there has yet to be an agreed-upon best method. However, several different approaches have been proposed. Many of these approaches either adapt procedures that have been successful in tree inference, such as maximum parsimony \cite{Jin2007, Park2010} and neighbor joining \cite{NeighborNet2004}, or building networks from a set of smaller inferred trees \cite{Baroni2005, Huber2011, RIATA2005, QuartetNet2014}. Many distanced-based methods have also been suggested \cite{Allman2022identifiability, Bordewich2018, Bordewich20182}, as well as methods that incorporate possible effects from incomplete lineage sorting using network extensions of the multispecies coalescence model \cite{Chifman2016, rabier2021inference, Solis2016, Wen2016, Yu2011, zhu2018bayesian}.  

There are known issues with several of the current methods (see, e.g. \cite{kong2022classes}, for a recent review). For example, when building networks from a set of 
small inferred trees, it is quite common for the method to overestimate the number of reticulation events \cite{Nakhleh2011}. Furthermore, it is known that methods where networks are constructed from 3-leaf trees may need to be more consistent \cite{Gambette2012}. For maximum parsimony, in \cite{Bryant2017} the authors show that the proposed methods return networks with too few or too many reticulation events depending on the maximum parsimony algorithm used. The methods based on the network multispecies coalescent model have had the most recent success,  with many implemented in  PhyloNet  \cite{Wen2018} for general use,  however, scalability remains an issue \cite{Hejase2016}. In addition, these methods often fail to identify 3-cycles and reticulation nodes in 4-cycles due to identifiability issues \cite{banos2019identifying, Solis2016}.

The method we propose is model-based, using network-based Markov models, and aims to address the issues above by using a combination of algebraic and statistical learning techniques to infer networks on small sets of taxa. We focus on the inference problem for 4-leaf level-one networks, that is, level-one \emph{quarnets}. It has already 
been shown that it is theoretically possible to construct level-one networks from their quarnets \cite{huber2018quarnet, VanIersel2014}
and more efficient puzzling techniques for reconstructing larger networks from smaller networks are quickly being developed \cite{huber2017reconstructing, huebler2019constructing}. Thus, the ability to accurately infer level-one quarnets provides the foundation
for reconstructing level-one networks of arbitrary size.

\section{background}
\label{sec:background}

We begin this section with an introduction
to the basic definitions and terminology for 
phylogenetic networks. We then introduce the 
particular phylogenetic network model that underlies
our method and the algebraic perspective on these networks models. Finally, we give a brief overview of support vector machines, the statistical learning approach that forms the basis of our method. 

\subsection{Level-one semi-directed networks}
\label{subsec:level1}


\begin{defn}
\label{def: network}
A \emph{rooted binary phylogenetic network}
$N$ on a set of leaves $[n]=\{1, \ldots, n\}$ is a rooted acyclic directed graph with no edges in parallel (i.e., no multiple edges) satisfying the following properties:
\begin{enumerate}
\item The root has out-degree two.
\item The only vertices with out-degree zero are the leaves, and each of these have in-degree one.
\item All other vertices either have in-degree one and out-degree two, or in-degree two and out-degree one.
\end{enumerate}
\end{defn}

The vertices of in-degree two in a phylogenetic network are referred to as 
\emph{reticulation vertices} and the \emph{level} of a phylogenetic network is the 
maximum number of reticulation vertices in a biconnected component of the network. Edges that are directed into reticulation vertices are referred to as \emph{reticulation edges}.
Given a binary phylogenetic network, if we undirect all non-reticulation edges of the network 
and suppress the root vertex, we obtain a \emph{phylogenetic semi-directed network}; such graphs are called \emph{semi-directed graphs} since some of the edges, in this case, the reticulation edges, are directed, while others are undirected. The set of phylogenetic semi-directed networks are exactly those leaf-labeled semi-directed graphs that can be obtained in this way. 
As an example, Figure \ref{fig: 4LeafRootedNetwork} shows
a 4-leaf rooted binary phylogenetic network alongside its associated
phylogenetic semi-directed network. The level of a phylogenetic semi-directed network is defined just as for a phylogenetic network as
are the reticulation vertices and edges.
These semi-directed networks are important for our purposes
since the location of the root of the phylogenetic network parameter 
is unidentifiable from the site-pattern probability distributions
produced by the models we consider \cite[Section 2.3]{gross2021distinguishing}.
Thus, all of the information about a Markov model on a binary 
phylogenetic network is contained in the associated phylogenetic semi-directed
network.

\begin{figure}[h]
\label{fig: semidirectedex}
\centering
\includegraphics[width = 7cm]{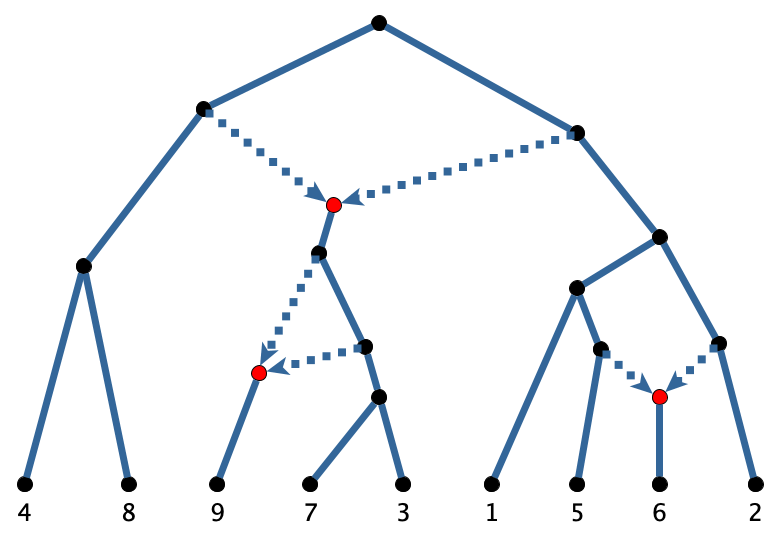}
\caption{A level-one 9-leaf rooted binary phylogenetic network with three reticulation vertices highlighted in red and six reticulation edges distinguished by dotted lines.}
\end{figure}

Figure \ref{fig: QuarnetPoset} represents all 4-leaf
level-one binary phylogenetic semi-directed networks. Beginning in the bottom row, we refer to these as \emph{trees}, \emph{3-cycle networks}, \emph{4-cycle networks}, and \emph{double-triangle networks}. In each of the 4-cycle networks,
the reticulation vertex is the large red vertex and the reticulation edges are
the two edges of the cycle directed into this vertex. For reasons of algebraic identifiability, which we discuss in Section \ref{sec: distinguishing invariants}, we do not show the reticulation edges on either the 3-cycle networks or double-triangle
networks. Hence, each 3-cycle network shown actually represents three different
phylogenetic semi-directed networks that can be obtained by specifying the reticulation vertex in the 3-cycle of the graph. Each double-triangle network represents eight different phylogenetic semi-directed networks obtained by choosing the reticulation vertex in each triangle. Note, choosing the two adjacent vertices in the triangles to be reticulation 
vertices does not result in a phylogenetic semi-directed network, since there is no root location compatible with the necessary edge orientations. We will refer to all network edges in a cycle as \emph{cycle edges}, and all edges adjacent to at least one cycle as \emph{cycle-adjacent edges}.

\subsection{Network Models of Sequence Evolution}
\label{subsec:networkmodels}

The method we suggest in this paper is based on the underlying network-based Markov model of sequence evolution described in \cite{Gross2017distinguishing, Nakhleh2011}. 
For a particular choice of parameters, a network-based Markov model returns a probability distribution on the $n$-tuples of DNA bases that may be observed at a particular site in the aligned DNA sequences of a set of $n$ taxa. Each of these \emph{site-pattern distributions} in the network-based model are weighted sums of site-pattern distributions belonging to tree-based phylogenetic models. 

In a tree-based phylogenetic model, evolution is modeled as a $k$-state Markov process 
proceeding along a rooted $n$-leaf phylogenetic tree $T$ with root $\rho$, where each vertex $v$ of $T$ is associated to a random variable $X_v$.  
In this paper, we will be concerned specifically with models of DNA sequence evolution, 
and so we let $k = 4$ and identify the states with the set of DNA bases $\{A, C, G, T\}$. Furthermore, since we will be concerned with \emph{quarnets}, i.e., 4-leaf phylogenetic networks or semi-directed networks, we will also assume $n=4$.

Let $\Delta^{d}$ be the $d$th dimensional probability simplex $\Delta^{d} := \{ p \in \mathbb R^{d+1} \ | \ \sum_{i=1}^{d+1}  p_i =1, \ p_i \geq 0 \text{ for } 1 \leq i \leq d+1\}$. A distribution in the tree-based Markov model associated with a tree $T$ is given by specifying a \emph{root distribution},  a vector $\pi \in \Delta^3$  defined by $P(X_\rho = i) = \pi_i$, and 
a Markov transition matrix $M^{(u,v)}$ with $M^{(u,v)}_{ij} = P(X_v = j | \ X_u = i)$ to each edge $(u,v)$ of $T$.
The transition matrices encode the probability of mutations
occurring along each edge of the tree. For phylogenetic analysis, we are particularly interested in
the states at the four leaves of $T$. 
These \emph{site-patterns} are the $4$-tuples of the DNA bases that we may 
observe in the aligned DNA sequences for a set of species.
To compute the probability of observing a particular site-pattern at the leaves, 
we marginalize over all possible states of the non-leaf vertices of $T$. In particular, let $\phi: V(T) \rightarrow \{A, C, G, T\}$ be an assignment
of states to the vertices of $T$; we can think of $\phi$ as a vector of length $|V(T)|$ and use $\phi_v$ to denote the state of $X_v$.  Let $\phi_{\mathcal{L}}$ be the restriction of $\phi$ to the leaves of $T$.  Then the probability of observing the 4-tuple $\omega \in \{A, C, G, T\}^4$ is

$$
\sum_{(\phi \ : \ \phi_{\mathcal{L}} = \omega)}\pi_{\phi_{\rho}}
\displaystyle \prod_{(u,v) \in E(T)} M^{(u,v)}_{\phi_u,\phi_v}.
$$

Thus, a four leaf tree $T$ defines a map 
$\psi_T: \Theta_T \rightarrow \Delta^{256 - 1}$ from the parameter space 
$\Theta$, which includes the parameters of the root distribution and the entries of the Markov
transition matrices, to the set of probability distributions on the $4^4=256$ possible
site-pattens that may be observed at the leaves of $T$. 
The \emph{model associated to T} is defined to be $\mathcal{M}_T := \im(\psi_T)$.  A key observation from algebraic statistics that will allow us to use algebraic methods is that the map $\psi_T$ is a polynomial map in the parameters of the model.

Similar to tree-based models, a phylogenetic network model on a 4-leaf phylogenetic
level-one network $N$ defines a map 
$\psi_N: \Theta_N \rightarrow \Delta^{256 - 1}$ from the parameter space
of the network model to the set of site-pattern distributions.
Again, similar to tree models, the parameter space 
includes a root distribution and a Markov
transition matrix associated to each edge of the network. 
However, for a phylogenetic network model with $m$ reticulation vertices $v_1, \ldots, v_m$, 
there are $m$ additional parameters $\alpha_{i} \in [0,1]$ for $1 \leq i \leq m$.  In the case of level-one quarnets, the number of reticulation vertices $m$ is at most two.

Each parameter $\alpha_{i}$ is arbitrarily associated to one of the reticulation 
edges $e^0_i$, directed into $v_i$. 
The pattern of inheritance at $v_i$ is directed through
the edge $e^0_i$ with probability $\alpha_{i}$ and through the other reticulation edge 
$e^1_i$ with probability $(1 - \alpha_{i}).$ 
Thus, to produce a site-pattern from the network model, 
for each $1 \leq i \leq m$ we independently select either 
$e^0_i$ with probability $\alpha_i$, or 
$e^1_i$ with probability $(1 - \alpha_{i})$, 
and remove the selected edge.
After removing one of each pair of reticulation edges in a level-one phylogenetic network,
the result is a $4$-leaf phylogenetic tree with the original leaf set.
A site-pattern probability distribution can be obtained from this tree-based model, as described above.
Therefore, the map $\psi_N$ can be described as a weighted sum of the
maps associated to the $2^m$ trees obtained by deleting one of
each pair of reticulation edges. The \emph{model associated to N} is defined to be $\mathcal{M}_N := \im(\psi_N)$.  Again, we note that the map $\psi_N$ is a polynomial map.

For tree-based and network-based Markov models in phylogenetics, it is quite common to simplify the model by adding constraints on the transition matrices. Such constraints reduce the dimension of the parameter space and the dimension of the images of $\psi_T$ and $\psi_N$. In this paper, we will assume the 4--state Jukes--Cantor model of 
DNA sequence evolution in which the root distribution is assumed to be
uniform and each Markov transition matrix has the form
$$
\begin{pmatrix}
1 - 3\beta & \beta & \beta & \beta \\
\beta & 1 - 3\beta & \beta & \beta \\
\beta & \beta & 1 - 3\beta & \beta  \\ 
\beta & \beta & \beta & 1 - 3\beta \\ 
\end{pmatrix}
$$
for some $\beta \in [0,1/4]$. 
The $\beta$ parameter in the Markov transition matrix along a particular edge
encodes the confounded effects of time and mutation rate along that edge.
Thus, branch lengths are often given in terms of \emph{expected number of substitutions per site}. These are the units used by the function {\tt{seqgen}} from the ${\tt R}$ package $\tt{phyclust}$, which we use to generate sequences. In these units, an edge of length $\ell$
corresponds to $\beta = \frac{1}{4}(1-  \exp(-\frac{4\ell}{3})).$

Because the Jukes--Cantor model is time-reversible, the root in either a
tree or network-based Jukes--Cantor model cannot be identified \cite{Felsenstein1981,Gross2017distinguishing, gross2021distinguishing}.
Consequently, for rooted networks $N_1$ and $N_2$, the models
$\mathcal{M}_{N_1}$ and $\mathcal{M}_{N_2}$ 
will be equal if $N_1$ and $N_2$
yield the same phylogenetic semi-directed network after unrooting.
Therefore, given data produced by a Jukes--Cantor phylogenetic network model, it is only possible to recover the  phylogenetic semi-directed network obtained by unrooting the network parameter \cite{Gross2017distinguishing}. For this reason, we define the models using phylogenetic semi-directed networks (as in Example \ref{ex: semidirected model}) and work only with these structures for the rest of the paper. 

\begin{ex}
\label{ex: semidirected model}
For the Jukes--Cantor phylogenetic model on the 4-leaf binary phylogenetic network depicted in Figure \ref{fig: 4LeafRootedNetwork}, the parameters of the model include the lengths of the edges and the two reticulation edge parameters (which must sum to one). As noted above, the root location of the phylogenetic network is not identifiable. 
Thus, we can equivalently construct the model by assigning edge lengths
and reticulation edge parameters to the semi-directed network shown in the same figure. The site-pattern probability distribution from the network will then be a weighted sum of the two site-pattern probability distributions coming from the tree-based Markov model with the edge lengths shown on the two trees at the right in the figure.

\begin{figure}[h]
\caption{The four structures in the figure below include, from left to right: a level-one 4-leaf rooted binary phylogenetic network, the associated 
phylogenetic semi-directed network, and the two unrooted trees were obtained by deleting reticulation edges in the semi-directed network.}
\centering
\includegraphics[width = 12cm]{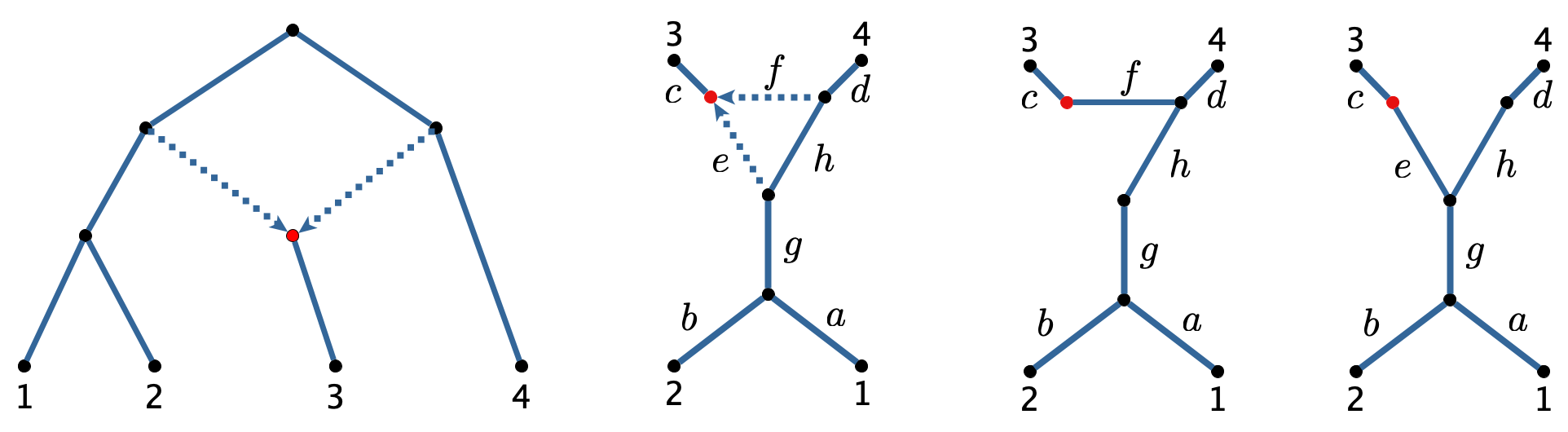}
\label{fig: 4LeafRootedNetwork}
\end{figure}
\end{ex}

\subsection{Phylogenetic Network Invariants}
\label{subsec:invariants}

There has been much work previously on identifying 
\emph{phylogenetic invariants} for Markov models of DNA sequence evolution
\cite{allman2007phylo}. 
The phylogenetic invariants for a tree-based Markov model on a tree $T$ are the 
polynomials that vanish on the model $\mathcal{M}_T$. 
That is, letting $[n]$ be the set of leaf labels for $T$ and letting $p_{i_1 i_2 \ldots i_n}=P(X_1 = i_1, \ldots, X_n = i_n)$, they are the polynomials contained in the ring
$$\mathbb{C}[p_{i_1 i_2 \ldots i_n}: i_1, i_2, \ldots, i_n \in \{A, C, G, T\}]$$
that evaluate to zero when the entries of any
site-pattern probability distribution from the model are substituted.
The set of all such polynomials that vanish on the model $\mathcal{M}_T$ is the ideal 
$$I_T \subseteq \mathbb{C}[p_{i_1 i_2 \ldots i_n}: i_1, i_2, \ldots, i_n \in \{A, C, G, T\}],$$
which is called the \emph{ideal of phylogenetic invariants} for $T$.
One motivation for computing these ideals is that they can be used to 
show that the models for different trees are not contained in one another. 
For example, showing  
$I_{T_1} \not \subseteq I_{T_2}$
proves the reverse non-containment for the models,
$\mathcal{M}_{T_2} \not \subseteq \mathcal{M}_{T_1}$.
This observation has been used to show that the tree parameter of
several tree-based Markov models is \emph{generically identifiable}.
That is, for a generic probability distribution coming from a tree-based
Markov model, it is possible to recover the tree parameter(s) of the model
(e.g., \cite{Allman, Allman2006, Rhodes2012}). 

In the same manner, as for trees, we can consider the ideal of phylogenetic invariants
for the network $N$ as the set of all polynomials that vanish on the model
$\mathcal{M}_N$. In \cite{Gross2017distinguishing} and 
\cite{gross2021distinguishing}, the ideals 
for several small networks were computed in order to show that the 
network parameter of certain network-based Markov-models is identifiable. 
This approach to establishing identifiability also suggests a method for 
phylogenetic inference. Specifically, suppose we have shown that for 
two networks $N_1$ and $N_2$ that
 $I_{N_1} \not \subseteq I_{N_2}$ and
$I_{N_2} \not \subseteq I_{N_1}$.
Then if the network parameter is generically identifiable, 
given a generic site-pattern probability distribution $p$ from either
$\mathcal{M}_{N_1}$ or $\mathcal{M}_{N_2}$, the distribution 
$p$ belongs to only one of these models. 
We can then substitute $p$ into a set of polynomials
that generate $I_{N_1} $ and 
a set of polynomials that generate $I_{N_2}$.
The result will be zero for every polynomial in the generating set of the 
ideal of the network with model containing $p$, and non-zero 
for at least one of the polynomials in the ideal of the network with model not containing $p$. 

Theoretically, this same principle should allow us to infer phylogenetic
networks from the observed site-pattern probability distributions coming from 
the aligned DNA sequences of a set of species. 
%
In practice, none of the phylogenetic invariants for any network 
are likely to evaluate to zero on observed data. 
This is not just because of the simplifying assumptions in our models.
Even if we simulate from a network-based Markov model to obtain an observed
site-pattern probability distribution, because the models are stochastic, we are likely
to get small non-zero values when we substitute the entries of this distribution into the 
phylogenetic network invariants.
Thus, using network invariants
for inferring phylogenetic networks requires
some means of interpreting the residuals, or deviations from zero, 
when observed site-pattern frequencies are substituted into the invariants.

One approach to doing this is to compute a score for each possible network
by adding up the absolute value of the residuals of the invariants in a generating
set for the ideal of the network. 
The inferred network would then be the one with the lowest score. 
This approach was used to infer phylogenetic trees in \cite{CF06, Rusinko2012}, 
though it is unlikely to be successful in our case. 
Those works considered only quartet trees which all have
the same unlabeled topology.
This allows for direct comparison of the computed scores
since the generating sets of invariants used are related by 
a permutation of the variables. 
However, in our case, we require invariants for quarnets with four distinct unlabeled quarnet topologies (the different rows in Figure \ref{fig: QuarnetPoset}). 
It is unclear exactly how one would compare scores for quarnets in different rows using different numbers of invariants of different degrees and with differing numbers of terms. 

A more refined strategy for addressing this problem would
be to apply statistical learning techniques to learn the patterns of residuals
from observed phylogenetic data. However, this is not possible
since there is a dearth of reliable labeled data from which to learn.
Yet another approach would be to derive expressions for the distribution of 
invariant residuals assuming a certain phylogenetic network model of evolution.
After all, for a fixed choice of model parameters, the invariants 
are polynomial functions of random variables.
However, this would require specifying a prior distribution
on the numerical parameters of the network models.
Moreover, the number of variables involved and the correlation
between them makes this infeasible.

For these reasons, we propose a method using support vector machines 
and random sampling in order to interpret the invariant residuals.
We begin by constructing a set of invariants $\mathcal{S}$ 
that distinguishes between all twenty-four level-one quarnets as shown in Figure \ref{fig: QuarnetPoset}. We then sample from a large region
of the numerical parameter space for each of the quarnet models.
The labeled sampled data is then transformed and substituted into $\mathcal{S}$,
and we train support vector machines on the invariant residuals.
The support vector machines can then be used to infer quarnets for observed
biological data. As previously noted, since all level-one phylogenetic networks 
can be constructed from their quarnets, this method can be paired with
any method for constructing level-one networks from quarnets to 
infer level-one networks of arbitrary size. 

\subsection{Support Vector Machines}
\label{sec: svm}

Our method relies on the construction of a support vector machine (SVM), a supervised learning model, for classifying the invariant residuals. We provide here a brief overview of SVMs adapted from \cite{James2017} and refer the reader there for more details. 

A linear support vector machine uses separating hyperplanes for classification. In the simplest case, the training data consists of observations in a Euclidean space each labeled as belonging to one of two classes. If the training data are separable, then there exists a hyperplane that perfectly separates the data so that the observations belonging to each class live on opposite sides of the hyperplane. In such a case, the separating hyperplane chosen is the \emph{maximal margin classifier}, which is the separating hyperplane that maximizes the distance to the nearest point in the training data. Once the hyperplane is determined, new observations can be easily classified by determining on which side of the hyperplane they reside. 

Of course, it is often not the case that the data used to train an SVM are perfectly separable. In these cases, the maximal margin classifier does not exist, and instead, we seek a \emph{soft margin classifier}. A soft margin classifier is again a hyperplane trained on the data; however, since the data are not separable, it will necessarily misclassify some observations in the training data. The soft margin classifier is determined by again choosing the hyperplane that best separates the training data according to some optimization criteria. Only now, when determining the optimal soft margin hyperplane, a cost is incurred for each misclassified observation and the total allowable cost must remain below a chosen threshold. For example, if the allowable cost were zero, then the soft margin classifier would only exist if the data were perfectly separable, and in that case it would be the maximal margin hyperplane. As is typical with statistical modeling, the cost parameter reflects a tradeoff between bias and variance and the optimal cost is typically determined through cross validation on the training set.

Two other issues distinguish most applications of SVMs from the simple case of two separable classes that we outlined above. The first issue is that even a soft margin classifier will only be effective at classifying new observations if the boundary between each pair of classes is approximately linear. However, it is easy to imagine applications where this is not the case. Consequently, SVMs are often constructed using \emph{kernels}, which transform the original observations, possibly by embedding them in a higher dimensional space. 
Non-linear decision boundaries in the original space can be represented as linear decision boundaries defined by hyperplanes in the transformed space. The second issue is that we may have observations belonging to one of several classes, rather than just two. There are a few ways to address this issue, one of which is the ``one-versus-one" approach, which we use below. In this approach, given data belonging to $m$ different classes, we construct $m \choose 2$ hyperplanes, one for each pair of classes. New observations are then classified by allowing each of these hyperplanes to ``vote," and the ultimate classification is the class that receives the most votes.

In our application, the observations are vectors of invariant residuals obtained by sampling from a network-based Markov model. Each observation has a label from the set $\{1, 2, \ldots, 24\}$ based on the equivalence class of the semi-directed network that produced the observation.
To build the SVM and classify points, we use the \R \ package {\tt{e1071}} \cite{e1071}. By default, this package uses a one-versus-one approach to classify new observations. We tune the cost parameter by trying several different values and evaluating the accuracy. 
Using residuals from a carefully constructed set of invariants, our data are already transformed in a way designed to theoretically separate the classes. Specifically, when comparing two quarnets, some of the coordinates correspond to distinguishing invariants for the two corresponding quarnets. We thus expect some of these coordinates to be near zero for one quarnet and non-zero for the other. Theoretically, it is possible that the distinguishing invariants will be near zero for one quarnet, and will lie above and below zero for the other in a way that makes a linear decision boundary a poor choice. However, this is not what we observe in practice, and typically the invariant residuals are near zero for the ``correct" quarnet and bounded away from zero for the other. Thus, we have found that a linear kernel is effective on these transformed coordinates, which we use in the simulations below.

While one could apply any of several different supervised learning methods to the classified invariant residuals, support vector machines offer some theoretical and practical advantages. 
First, on the theoretical side, support vector machines have an intrinsic geometric interpretation. 
Since we are viewing the phylogenetic network inference problem from an algebro-geometic vantage point, this may prove useful for further inverstigation. For example, the coefficients of the SVM hyperplanes may offer clues as to the relative importance of invariants for separating different models. From a practical perspective, support vector machines performed better in our initial explorations than other supervised learning methods (e.g., decision trees, $k$-nearest neighbors).  
Moreover, using results from \cite{lin2007note},  \cite{platt1999probabilistic}, and \cite{wu2003probability}, one can modify this method to return probabilities for each class rather than a classification. This may be useful for developing methods for constructing larger networks from quarnets using a weighted quarnet scheme similar to those for trees \cite{Strimmer1996, Ranwez2001}.

\section{Constructing A Distinguishing Set of Phylogenetic Network Invariants for 
Jukes--Cantor Quarnets}
\label{sec:quarnetinvariants}

As noted previously, our analysis will focus on inferring 
level-one quarnets, or 4-leaf semi-directed networks, 
since these can be used to build larger networks. 
We will also assume that the underlying DNA substitution process
is the 4-state Jukes--Cantor model. 
In order to apply the invariants
based inference method described in the previous
section, we first need the following definition. 

\begin{defn}
\label{def: distinguishing set}
Let $\mathcal N$ be a set of phylogenetic networks and $\mathcal{M}$ a 
phylogenetic model. A set $\mathcal{S}$ is a 
\emph{distinguishing set of invariants for $\mathcal N$ under $\mathcal{M}$},
if for all $N_1, N_2 \in \mathcal N$, there exists
a polynomial invariant $f$ in the vanishing ideal of 
$I_{N_1}$ or $I_{N_2}$, but not both. 
\end{defn}

\noindent Thus, our first step is to construct a set of
invariants $\mathcal{S}$ that is a distinguishing set
of invariants for the set of quarnets under the Jukes--Cantor model.  

\subsection{Generating sets of network ideals and distinguishing invariants}
\label{sec: distinguishing invariants}

Generating sets for the vanishing ideals of all level-one quarnets with a 
single cycle are known from \cite{Gross2017distinguishing} and those of the three level-one quarnets with two reticulation
vertices (the double-triangle networks) from 
\cite{gross2021distinguishing}. 
The computations for these ideals are
contained in the supplemental materials 
of those works.

The ideals are actually computed in a set of transformed coordinates.
For quarnets under the Jukes--Cantor model,
the ring of transformed coordinates is the ring of \emph{$q$-coordinates}, also referred to as \emph{Fourier coordinates},
$$\mathbb{C}[q_{i_1 i_2 i_3 i_4}: (i_1, i_2, i_3, i_4) \in \{A, C, G, T\}^4].$$
These $q$-coordinates are computed from the \emph{probability 
coordinates} as follows.
Letting $\chi$ be the matrix
$$
\begin{pmatrix}
1 & 1 & 1 & 1 \\
1 & -1 & 1 & -1 \\
1 & 1 & -1 & -1 \\
1 & -1 & -1 & 1 \\
\end{pmatrix}$$ 
with rows and columns
indexed by $A, C, G$ and $T$, 
\begin{equation}
\label{eq: transform}
q_{i_1i_2i_3i_4} = 
\displaystyle \sum
\chi_{i_1, j_1}
\chi_{i_2, j_2}
\chi_{i_3, j_3}
\chi_{i_4, j_4}
p_{j_1j_2j_3j_4}
\end{equation}
where the index is over all
$(j_1, j_2, j_3, j_4) \in  \{A, C, G, T\}^4.$
This linear change of coordinates, called the \emph{Fourier transform},
was introduced in \cite{Evans1993} and is common in phylogenetics
because it simplifies the description of the models. 
For example, for tree-based models, after applying the Fourier
transform to both the domain and image spaces of the map
$\psi_T$, the ideal $I_T$ is generated by binomials in the $q$-coordinates.
The details of the transform are not 
particularly relevant to our analysis, so we refer the interested 
reader instead
to \cite{Evans1993, Sturmfels2005}. For our purposes, the most important point is
that since our invariants are expressed in the $q$-coordinates,
we will need to first transform the observed site-pattern frequencies
using (\ref{eq: transform}).

Due to the symmetry of the Jukes--Cantor model, we will not
need to compute the entire lvector of 256 $q$-coordinates.
This is because there are a number of site-patterns that are predicted to appear
with the same frequency regardless of the network or tree parameter of the model. 
For example, because the rate of substitution between all sites
is the same under the Jukes--Cantor model, 
site-pattern probability distributions from every network
will satisfy $p_{ACTA} = p_{GTAG}$. Put another way, this 
is a linear invariant in the probability coordinates 
that is contained in the ideal of every quarnet.
This symmetry in the probability coordinates simplifies the resulting
$q$-coordinates. For example, 128 of the $q$-coordinates are identically
zero for every site-pattern probability distribution contained in a
Jukes--Cantor model.
Moreover, many of the other non-zero coordinates are identical. 
Working modulo these linear invariants, we can express the ideals for
each of the quarnets in the ring containing only the following 15 variables,

\begin{align}
\mathbb{C}[
&q_{AAAA}, 
q_{AACC}, 
q_{ACAC}, 
q_{ACCA},
q_{ACGT}, \nonumber \\
&q_{CAAC}, 
q_{CACA},
q_{CAGT},
q_{CCAA}, 
q_{CCCC},\label{eq: qring}  \\ 
&q_{CGAT},
q_{CGCG},
q_{CGTA},
q_{CCGG},
q_{CGGC}   \nonumber
].
\end{align}

The equivalence classes of these variables are listed in the catalogue of Small Phylogenetic Trees \cite{smalltrees} under the Jukes-Cantor model for a 4-leaf unrooted tree. 
Note that the catalogue assumes the additional symmetries of a particular tree and so the equivalence classes of
$q_{CGCG}$ and $q_{CGGC}$ as well as those of
$q_{CCCC}$ and $q_{CCGG}$ are combined.

Computing the vanishing ideals for each level-one quarnet 
under the Jukes--Cantor model reveals an obstacle
to computing a set of distinguishing invariants. First,
as observed in \cite{Gross2017distinguishing}, 
there are many level-one quarnets with identical ideals. 
Specifically, two 3-cycle networks
or two double-triangle networks with the same undirected skeleton
have the same vanishing ideal under the Jukes--Cantor model.
 For example, consider the 4-leaf rooted binary phylogenetic network pictured in Figure \ref{fig: 4LeafRootedNetwork}. If we were to switch the leaf labels $3$ and $4$ in this network, the
 undirected skeleton of the associated phylogenetic semi-directed remains unchanged. The same is true if we instead switch the leaf vertex labeled by $3$ and the cherry labeled by $(12)$ in the original network. 
 
\begin{figure}
    \centering
    \includegraphics[width = 12cm]{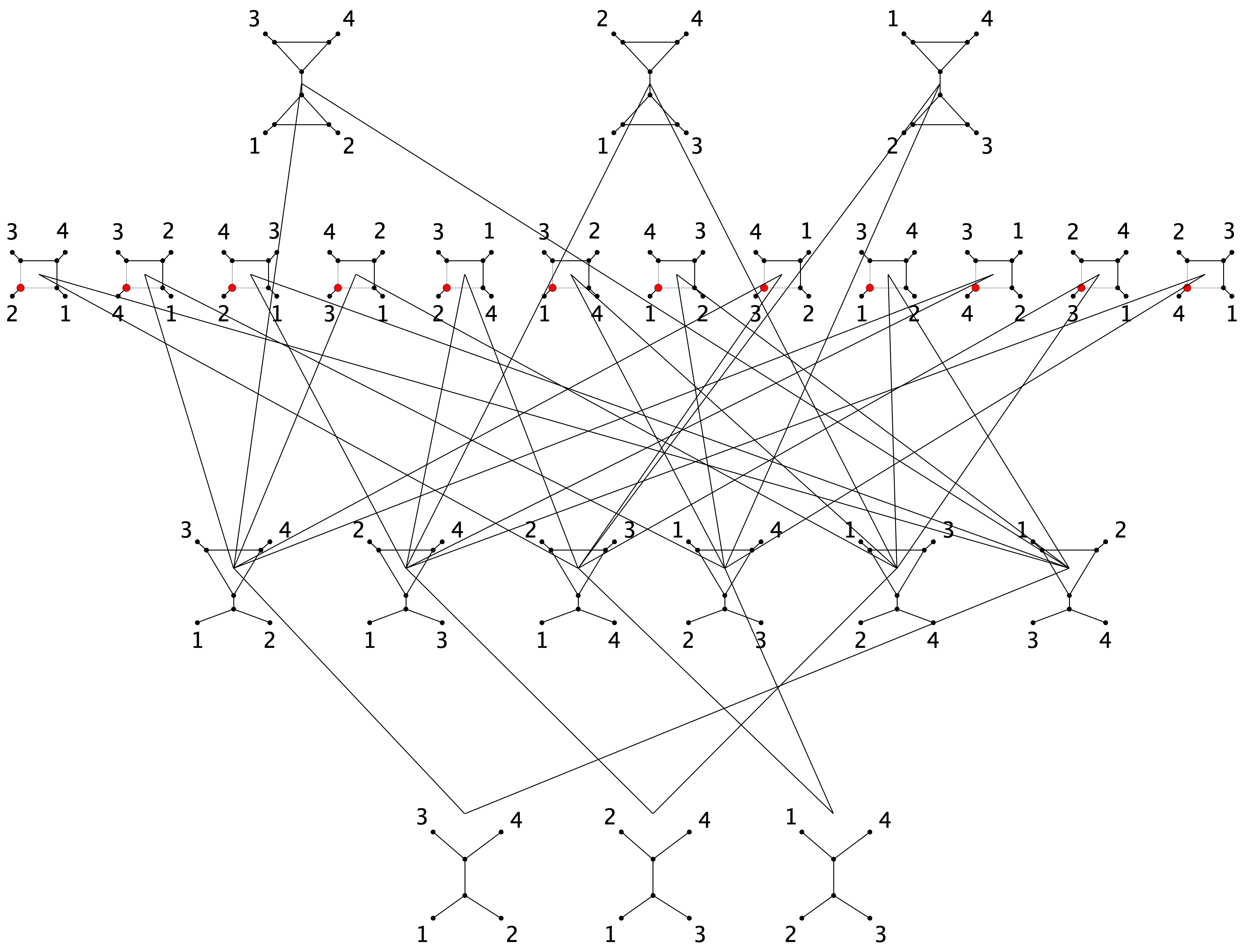}
    \caption{The poset of algebraically identifiable 4-leaf level-one phylogenetic semi-directed networks. A line between two quarnets indicates that the ideal of the quarnet above is contained in the ideal of the quarnet below
    }
    \label{fig: QuarnetPoset}
\end{figure}

 It may be that the Markov models built on these three distinct phylogenetic semi-directed networks are in fact identical. It is also possible that they are distinguishable in some way, for example, they may satisfy certain inequalities or non-polynomial invariants. For these
 reasons, when we generate samples for the 3-cycles and double-triangles
 networks, we randomly choose one of the valid orientations for the reticulation edges. 
 Still, since our method is based on theoretical results with polynomial invariants, we will only ever attempt to infer undirected 3-cycles.
Thus, from this point forward, when we refer to quarnets, we refer
to the 24 distinct topological structures depicted in 
Figure \ref{fig: QuarnetPoset}.

We also note that even among the quarnets depicted in Figure \ref{fig: QuarnetPoset}, 
the ideals of some quarnets are properly contained 
in the ideals of others. This containment is encoded 
in the figure, where a line between two quarnets
indicates that the ideal of the quarnet above is contained in
the ideal of the quarnet below. Thus, it is impossible
to form a distinguishing set that contains
an invariant that belongs to the vanishing ideal
$I_{N_1}$ but not to $I_{N_2}$ for all quarnets $N_1$ and $N_2$.

Recalling that the containment of ideals is reverse to the
containment of models, some of the above results 
come as no surprise. 
For example, it is clear that adding a reticulation edge to a tree results
in a model on a 3-cycle network that must contain the tree model. 
This follows since any distribution from the tree model can be obtained from the
3-cycle network model by setting one of the reticulation edge parameters to 0 and the other to 1.
Likewise, we would expect the models of the 3-cycle networks
to be contained in the models of the double-triangle networks formed
by adding an extra reticulation edge. However, it is perhaps
more surprising that the models of the 3-cycle networks
are contained in models for 4-cycle networks. 
This is not the case for the Kimura 2-parameter or Kimura 3-parameter model,
so this appears to be a feature of the Jukes--Cantor model rather
than a feature of the quarnets \cite{gross2021distinguishing}.

This is the reason that our definition of a distinguishing set of invariants for $N$ requires only
that for every pair of quarnets, $N_1$ and $N_2$ in $N$, the distinguishing set contains
an invariant that belongs to either $I_{N_1}$ or $I_{N_2}$, but not both. 
For example, consider the triangle quarnet $N_4$ and the 4-leaf tree $T_1$ with
$I_{N_4} \subset I_{T_1}$. 
If $f \in I_{T_1} \setminus I_{N_4}$,
then for a generic probability distribution $p$ contained in 
either $\mathcal{M}_{N_4}$ or $\mathcal{M}_{T_1}$,
$f(p) = 0$ if $p \in  \mathcal{M}_{T_1}$ and 
$f(p) \not = 0$ if $p \in  \mathcal{M}_{N_4}$.
Indeed, this is the basis of using invariants for inference.

\begin{ex}
\label{ex: varyinggamma}

The graph in Figure \ref{fig: LoopsVaryingGamma} shows the residuals of two invariants contained
in $I_{T_1} \setminus I_{N_4}$. 
Each ``loop" corresponds to a network obtained by randomly assigning edge lengths (between $0.1$ and $0.2$) to the 4-leaf semi-directed network depicted in Figure 
\ref{fig: 4LeafRootedNetwork}. 
Each point represents the $q$-coordinates of a theoretical site-pattern probability distribution
in the model $\mathcal{M}_{N_4}$ evaluated at the two invariants as the reticulation edge parameter $\gamma$ on the edge labeled by $f$ goes from 0 and 1. Hence, the loops start and end at the origin, since site-pattern probability distributions in the model 
$\mathcal{M}_{N_4}$ corresponding to reticulation edge parameters of 0 or 1 are also contained in the model $\mathcal{M}_{T_1}.$

\begin{figure}[h]
\caption{Plot of two invariant residuals in $I_{T_1} \setminus I_{N_4}$ as gamma varies for 
ten randomly chosen, fixed assignments of edge lengths to $N_4$.}
\label{fig: LoopsVaryingGamma}
\centering
\includegraphics[width = 8cm]{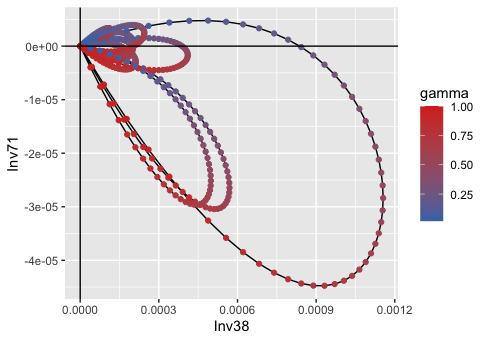}
\end{figure}

\end{ex}

\begin{ex}
\label{ex: nestedsubsets}

The data shown below represent samples of $10^6$ sites from 
100 randomly chosen samples in our training set, 50 each from the models $\mathcal{M}_{T_1}$
and $\mathcal{M}_{N_4}$.  In each of the plots, the plotted points correspond
to the residuals of two different invariants and the colored
regions correspond to the decision boundary of the
SVM trained on the plotted data. 

The invariant residuals shown in the first plot are 
from two invariants in $\mathcal{S}$
that distinguish these quarnets
(i.e., they belong to 
${I}_{T_1} \setminus {I}_{N_4}$). 
The invariants in the second plot do not distinguish these quarnets since they each vanish on both models 
(i.e., they belong to $I_{T_1} \cap I_{N_4}$).

\begin{figure}
\centering
\begin{subfigure}{.5\textwidth}
  \centering
  \includegraphics[width=7cm]{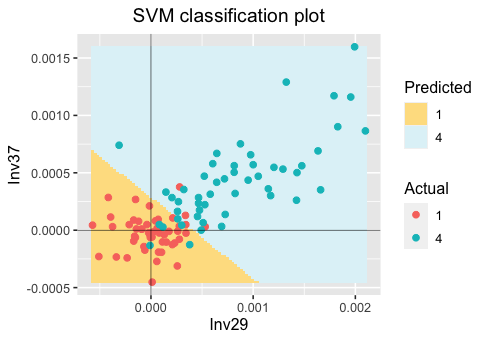}
  \caption{Invariants contained in ${I}_{T_1} \setminus {I}_{N_4}$.}
  \label{fig: invariants_distinguishing}
\end{subfigure}%
\begin{subfigure}{.5\textwidth}
  \centering
  \includegraphics[width=7cm]{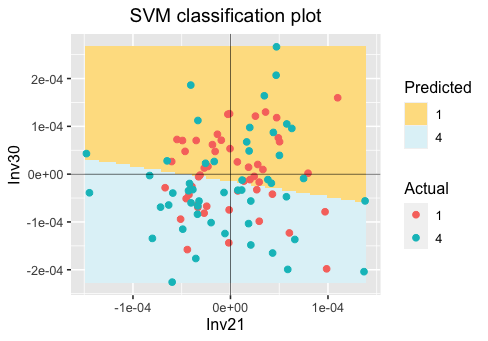}
  \caption{Invariants contained in
  $I_{T_1} \cap I_{N_4}$.}
  \label{fig: invariants_both}
\end{subfigure}
\caption{SVM regions for trained on the residuals of two invariants for a subset of the training data sampled from $\mathcal{M}_{T_1}$ and $\mathcal{M}_{N_4}$.}
\label{fig: svmregions}
\end{figure}

The theoretical basis for our method is depicted in the first plot.
The distinguishing invariants are much better able to distinguish between the two quarnets when compared to the invariants contained in both ideals.
When each of these SVMs are applied to the samples in the test set corresponding to $T_1$ and $N_4$, 
the accuracy of the model trained on distinguishing invariants
is 87.75\% compared to just 45.25\% for the other model.
Though not shown, we also applied the same method 
with two invariants in $\mathcal{S}$ that do not belong to either ideal.
Even though there is no theoretical basis for these invariants to perform well,
it would not be surprising if some polynomial transformations of the data
had some power to distinguish between quarnets.
In this case, the accuracy was 54.75\%--better than the two invariants contained in 
both ideals but still less than those that theoretically distinguish the network.

Different pairs of
invariants may behave differently, but to illustrate the principle, we chose pairs of invariants of low degree (so that the residuals were relatively large) that were uncorrelated (so that they did not contain the same information).
To do this, we converted $\mathcal{S}$ to a list and selected the first five low-degree invariants of each type from $\mathcal{S}$. 
For each type, we then selected the pair of invariants with the lowest absolute correlation of the residuals in the training set.
\end{ex}

\subsection{Permutation Invariance}
\label{sec:symmetric}

So far, we have specified only that the set $\mathcal{S}$ be a distinguishing
set. However, it is an open question in phylogenetics as to which invariants
are the most effective framework for inferring phylogenies.
One way to construct a distinguishing set is to find a generating set of each quarnet ideal and then take the union of these sets of polynomial invariants.
This still leaves many possible options for the distinguishing set since the generating 
set of an ideal is not unique. 
The computer algebra systems used to compute vanishing ideals return generating sets that satisfy nice mathematical properties, such as being minimal generating sets or being Gr\"{o}bner bases with respect to some term orders.
These properties do not, however, have a clear biological interpretation and may not be well suited for distinguishing phylogenetic models.

A desirable property of any phylogenetic inference algorithm is that it is invariant under permutations of the data. That is, if the algorithm returns network $N$ for input $p$, then 
it should return network $\sigma(N)$ for input $\sigma(p)$.
As an example, suppose we input the set of aligned sequences for taxa $ABCD$ in that order into a phylogenetic inference algorithm and that the algorithm returns $N_{10}$, the 4-cycle network
at the far left in Figure \ref{fig: QuarnetPoset}. 
This indicates that 
taxon $B$ is a hybrid of taxa $A$ and $C$. Now suppose that instead we input the aligned sequences in the order $ACDB$.
Since the relationship between the taxa has not changed, we expect the algorithm to return $N_{21}$, indicating that the fourth taxon ($B$) is a hybrid of the first and second taxa ($A$ and $C$ in the new ordering). 

This issue for invariants-based algorithms was first identified and addressed in the case of tree invariants in \cite{Rusinko2012}. 
In order to ensure that our algorithm is also invariant under permutations of the data, we adopt a similar approach and construct $\mathcal{S}$ itself to be \emph{permutation invariant} as we define below. 
Before we define this concept more formally, we need two small pieces of notation. One, for $\sigma \in S_n$ and a variable $q_{i_1 i_2 \ldots i_n}$, 
$\sigma(q_{i_1 i_2 \ldots i_n}) = 
q_{\sigma(i_1 i_2 \ldots i_n)}$. And two, 
for a polynomial $f \in 
\mathbb{C}[q_{i_1 i_2 \ldots i_n}: i_1, i_2, \ldots, i_n \in \{A, C, G, T\}]$, $\sigma(f)$ is the polynomial that results
from applying $\sigma$ to every variable in $f$.
As an example, let 
$$f = q_{ACCA}q_{CAGT} - q_{CGTA}q_{CACA}$$ and let
$\sigma$ be the transposition $(13) \in S_4$.
Then 
$$\sigma(f) = q_{CCAA}q_{GACT} - q_{TGCA}q_{CACA}.$$

\begin{defn}
\label{defn: permutation invariant}
A set of polynomials
$$\mathcal{S} = \{f_1,f_2,\cdots,f_l\}
\subseteq
\mathbb{C}[q_{i_1 i_2 \ldots i_n}: i_1, i_2, \ldots, i_n \in \{A, C, G, T\}]$$
is \emph{permutation invariant} if for any permutation 
$\sigma$ in $S_n$ and
any $1 \leq j \leq l$, 
$\sigma(f) \in \mathcal{S}$.

\end{defn}

In order to construct a permutation invariant set of distinguishing invariants we first let $\mathcal{S}$ be the union of generating sets for the ideals of the quarnets $T_1$, $N_4$, $N_{10}$, and $N_{22}$. That is, we include in $\mathcal{S}$ polynomials sufficient to generate the ideal for one quarnet from each row of Figure \ref{fig: QuarnetPoset}. We then apply a permutation $\sigma \in S_4$ to all of the polynomials in $\mathcal{S}$, and add these polynomials to $\mathcal{S}$. We repeat this process until $\mathcal{S}$ is permutation invariant (which occurs after applying just the six transpositions in $S_4$). 
Note that we still embed our distinguishing set in the ring of $q$-coordinate equivalence classes from (\ref{eq: qring}). 
For example, if we determine
$\sigma(f) = q_{CCAA}q_{GACT} - q_{TGCA}q_{CACA}$ 
should be in $\mathcal{S}$, then we replace
$q_{GACT}$ and $q_{TGCA}$ by their equivalence class
representatives and add the polynomial  
$q_{CCAA}q_{CAGT} - q_{CGTA}q_{CACA}$ to 
$\mathcal{S}.$

The set $\mathcal{S}$ we construct contains polynomials sufficient to generate the ideal for every quarnet in Figure \ref{fig: QuarnetPoset}. This is because applying a permutation to a generating set for the ideal of a quarnet produces a generating set for the ideal of another quarnet in the same row. 
This implies that $\mathcal{S}$ is a distinguishing set.
We have already shown that if $N$ and $N'$ are any two of the 24 quarnets, 
then without loss of generality we may assume that $I_{N} \not \subseteq I_{N'}$.
It follows then that any generating set of $I_{N'}$ contains a 
polynomial $f \in  I_{N'}\setminus I_{N}$, and so $\mathcal{S}$ is a permutation invariant distinguishing set as desired. The set $\mathcal{S}$ contains 1126 polynomials and is available in the supplemental materials in a text file which can be read into Macaulay2.

Constructing such a large distinguishing set increases the run time of our algorithm, particular in the construction of the support vector machines. 
However given the theoretical and practical implications of potentially returning different answers to an inference problem given equivalent input sets, we deemed the increase in run time from insisting $\mathcal{S}$ be
permutation invariant is worth the cost.
Note that $\mathcal{S}$ is 
not the smallest permutation invariant distinguishing set we could construct. For one, we could reduce the size of $\mathcal{S}$ by beginning with a 
\emph{minimal} generating set for the ideals $T_1$, $N_4$, $N_{10}$, and $N_{22}$. Moreover, it is not even strictly necessary for $\mathcal{S}$ to contain generators of every ideal in order for it to be a distinguishing set.
For example, we could construct a distinguishing set that contains just a few invariants (at most $24 \choose 2$),
and then expand this set until it is symmetric.
However, previous results have shown that some invariants perform much better than others at distinguishing between phylogenetic structures from data \cite{Casanellas2011, casanellas2015low}.
A priori, it is unclear which invariants will perform best, and it may be that certain invariants perform better than others at distinguishing between quarnets over different regions of parameter space. 
For these reasons, we chose to include a large number of invariants in $\mathcal{S}$ and to determine which are important through statistical learning and the construction of the support vector hyperplanes.
It is an interesting question though to determine the smallest permutation invariant distinguishing set for a given collection of networks and how to systematically construct such a set.

\section{Quarnet Network Reconstruction using Support Vector Machines (QNR-SVM). }
\label{sec:algorithm}

We call our invariants-based algorithm for inferring phylogenetic networks 
Quarnet Network Reconstruction using Support Vector Machines (QNR-SVM). 
The algorithm takes as input the aligned DNA sequences for a set of four taxa,
and then uses support vector classifiers to classify the input
data as belonging to one of the 24 quarnet models. 
The algorithm is implemented in \R, and the 
output is the quarnet associated to this model.
In this section, we describe this algorithm in further detail by 
first describing the training data and
the process used to construct the support vector classifiers.

\subsection{Training the Support Vector Machine}
\label{sec: training}

In order to construct a point in our training data set, we begin by sampling 
a site-pattern probability distribution from a quarnet model.
For a fixed quarnet $N$, a site-pattern probability distribution $p \in \mathcal{M}_N$ is determined
by the numerical parameters of the model, which
include the edge lengths of the quarnet, and for all quarnets that are not trees,
the reticulation edge parameters. 
For a fixed quarnet, we select cycle and cycle-adjacent edge lengths
uniformly at random from the closed interval $[0.05, 0.2]$
 and we select all other edge lengths uniformly at random from $[0.05,0.4]$. We select each of the reticulation edge parameters
uniformly at random from the closed interval
$[0.25, 0.75]$. We use different intervals for cycle and cycle-adjacent edges to ensure that sampling from a tree (or 3-cycle) is equivalent to sampling from a 3-cycle (or double-triangle) with a reticulation edge parameter set to 0. This way the distributions for each nested chain of quarnets lie roughly in the same region of probability space. 
If we did not use different intervals, then samples from a quarnet with a reticulation would display on average greater distances between taxa than the same quarnet without that reticulation; thus we run the risk of the SVM classifying based on this feature of the data, rather than on the underlying quarnet. 

Once the choice of numerical parameters is made, we sample a DNA sequence alignment from the model consisting
of $10^6$ sites for each taxon using the function {\tt{gen.seq.HKY}} (which calls the function {\tt{seqgen}})
from the  \R \   package  {\tt{phyclust}} \cite{phyclust}. As noted previously, a branch length $\ell$ in 
this function corresponds to a value of $\beta = \frac{1}{4}(1-  \exp(-\frac{4\ell}{3}))$ 
 in the Jukes-Cantor transition matrix on an edge.

We then convert this alignment to a length 256 empirical site-pattern probability distribution. Next, we average over Jukes-Cantor equivalence classes so that entries in the same equivalence class are equal. 
Finally, the resulting distribution is converted to $q$-coordinates using
Equation \ref{eq: transform} and these 15 entries are substituted into the invariants in $\mathcal{S}$ (with a fixed ordering).
The result is a vector of length 1126 which is labeled by the 
quarnet that produced the data. In the supplemental files,
we number the quarnets in Figure \ref{fig: QuarnetPoset}
from left to right and bottom to top so that the trees are numbered 1-3, 
the 3-cycle networks 4-9, the 4-cycle networks 10-21, and the double-triangle
networks 22-24.

Instead of repeating this process for each quarnet, 
we fix one representative of the four unlabeled quarnet topologies.
After the aligned DNA sequences are generated, we then permute
the sequences by each of the 24 permutations in $S_4$.
The resulting sequences are then converted into length 1126 vectors
as described and labeled with the quarnet obtained after applying 
the same permutation to the leaf labels. 
As a result, the entire set of training data is
also invariant under permutation of the $q$-coordinate indices, 
a condition that ensures our algorithm is permutation invariant.

Each independent sample from each of the four unlabeled quarnet topologies 
results in $|S_{4}| = 24$ observations added to our training set.
One important thing to note is the difference in the number of label permutations 
that fix each quarnet. 
For example, there are eight label permutations that fix a quarnet tree, but only
two that fix a 4-cycle. 
Therefore, a single independent sample from a quarnet tree, once permuted, will generate
eight observations labeled by each the three quarnet trees, whereas a single sample from
a 4-cycle will generate only two observations labeled by each 4-cycle.
In order to avoid a class imbalance, which can bias an SVM model towards the majority class 
\cite{Batuwita2013}, we sample so that there are 9600 observations labeled by each quarnet.
While this balances the number of observations labeled by each quarnet, the number of independent samples
corresponding to each quarnet will be different. For example, the observations corresponding to a 4-cycle
represent four times as many independent samples as those corresponding to a tree.
It is unclear exactly how this affects the construction of the support vector classifiers, but is a necessary
byproduct of our insistence on a training set with balanced classes that is invariant under permutation.

After the training data is constructed, we use the function 
{\tt{svm}} from the \R \ package {\tt{e1071}}  to construct our model.
We use a linear kernel and tune the {\tt{cost}} parameter of the 
soft-margin classifier by building several models with different {\tt{cost}} parameters (0.0001, 0.001, 0.01, 0.1, 1, 10) and comparing the returned accuracy of each model on an independent test set. By default, {\tt{svm}} scales each column, which allows us to meaningfully compare
the residuals from invariants of different degrees.
Also by default, the function uses a one-versus-one approach which we 
describe in Section \ref{sec: svm}. With $230400 =9600 \times 24$ training points, construction of the support vector classifiers takes a total of 21.4 hours on a iMac Pro with a 2.5 GHz Intel Xeon W processor. The resulting SVM model is available in the supplementary material.



\subsection{The QNR-SVM Algorithm and Simulation Results}
\label{sec:simstudy}

Almost all of the computational cost of the QNR-SVM 
algorithm comes from the generation of the training data
and the construction of the support vector classifiers,
which both need only be performed once. 
Assuming this has been done, applying the algorithm
is simply a matter of transforming the input data appropriately
for use in the SVM model. We summarize each step of the process here. These steps are implemented in 
 the supplementary files ({\tt SamplingFunctions.R} and {\tt TrainingAndTesting.R}).

\begin{itemize}
\item {\bf Step 1:} Given the aligned DNA sequences for four taxa, compute the length 256 site-pattern frequency vector $p$.
\item{\bf Step 2:} Replace each coordinate of $p$ with the average count for its Jukes-Cantor
equivalence class.
\item{\bf Step 3:} Use the discrete Fourier transform to convert the vector $p$ from probability coordinates to $q$-coordinates.
\item{\bf Step 4:} Construct the residual vector $r = (f_1(q),\cdots, f_{1126}(q))$ for $f_{i} \in \mathcal{S}$.
\item{\bf Step 5:}  Use the previously built SVM classifier to classify $r$ as belonging to one of the 24 network models. 
\end{itemize}

With an SVM classifier already in hand, the five steps listed above complete rather quickly.
For example, for an alignment of $10^6$ sites, the five steps complete in 22.21 seconds with Step 1, converting the aligned sequences into a site-pattern frequency vector, taking the most time (21.06 seconds), and Step 5, the classification step taking only 1.35 seconds. The speed that a quarnet is recovered is one of the benefits of this method, however, as with other methods, we expect scalability to remain an issue since the number of quarnets to classify grows combinatorially with the number of leaves.

We first demonstrate the performance of the QNR-SVM algorithm on model generated data.
This at least offers evidence
that the invariants-based SVM is able to learn general features of the network models that are not specific to the training set.
In Section \ref{subsec:Hybrid}, we apply the algorithm to a biological data set consisting of primate data. The results are largely in accordance with previously published results, suggesting the algorithm has practical applicability beyond model simulated data.

%
%
%
%
%

\begin{figure}
\begin{center}
\includegraphics[width = 14cm, height = 12cm]{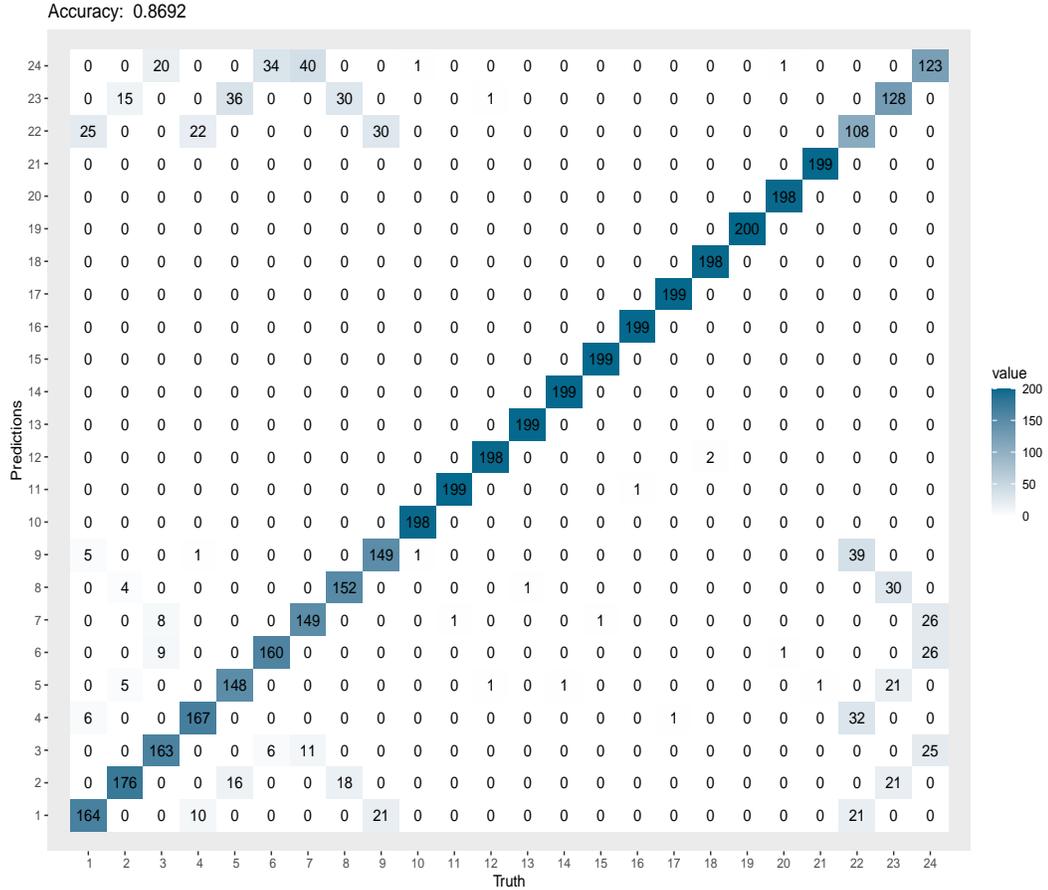}
\end{center}
\caption{Confusion matrix for 4800 test observations, 200 for each quarnet. The true quarnet labels for the simulated observations are listed along the bottom of the plot, while the predicted quarnet labels are listed along the left side of the plot. A 200 in a diagonal entry means that 100\% of the points sampled from the corresponding quarnet were classified correctly.  }
\label{fig:confusion-matrix}
\end{figure}

Figure \ref{fig:confusion-matrix} displays the
predictions results using QNR-SVM. The classifier was trained according to the specifications detailed in Section \ref{sec: training}. The test set consists of 4800 observations that result from independently sampling 200 observations from each of the 24 quarnets.
Each independent sample was obtained by generating an
alignment of $10^6$ sites for each taxon from one of the quarnet models following the same branch length sampling regime 
used for the training set.
   
The confusion matrix in Figure \ref{fig:confusion-matrix} reveals the strengths and weaknesses of the method. The overall accuracy on our test set is 86.92\%. The method performs particularly well with 4-cycles---more than 99\% of the observations generated by a 4-cycle quarnet are correctly classified. Likewise, very few observations generated by one of the other quarnets are misclassified as having been generated by a 4-cycle. The method has more difficulty classifying observations sampled from trees, 3-cycles, and double-triangles. However, a closer look reveals that the misclassifications follow a consistent pattern.
In almost all cases, the misclassified observations are incorrectly assigned to a submodel or supermodel of the correct model. Put another way, the observations are assigned to a quarnet that results from adding a reticulation edge to or deleting a reticulation edge from the correct quarnet. Indeed, this phenomenon explains the symmetric patterns of misclassified quarnets we see off the diagonal in the confusion matrix. This pattern is perhaps not all that surprising. For example, correctly classifying these observations requires in some cases determining if an observation was sampled from a 3-cycle with a reticulation edge parameter near 0 or 1, or a tree that is essentially that same 3-cycle with reticulation edge parameter exactly 0 or 1.  
  More precisely, we see that, in this test set, all but one, i.e. $99.95 \%$, of observations from trees, 3-cycles, and double-triangles are either correctly classified, or classified as a quarnet with an additional or missing reticulation.

  Due to the underlying nested geometry of the models, we expected the classifier to have difficulty with triangles, and, as noted in the introduction, it is known that triangles are hard to identify in many network inference methods. In the experiment reported here, we attempted to avoid the degeneration issue that happens as reticulation edge parameters approach 0 and 1 by bounding these parameters away from 0 and 1 when generating
our data for the training set and test set. As a consequence, there is some built in preference for the method to default to 
a quarnet without a specific reticulation unless there is strong evidence for it,
which is line with the general practice of inferring the most parsimonious model
in order to explain the data. However, as we see from the confusion matrix, there is room to strengthen this preference. One way this can be done is by putting a threshold on the votes or outputted probabilities. The outputted class probabilities for a SVM in the ${\tt e1071}$ package are obtained according to the methods in \cite{wu2003probability}, where pairwise probabilities are estimated by fitting a logistic distribution to the decision values and then the class probabilities are obtained by reducing the problem to a quadratic optimization problem. For example, for the test points corresponding to the 3-cycle $N_4$ classified in Figure 6, the outputted probabilities when the network was classified correctly displayed a strong signal with probabilities in the $80 - 90 \%$ range for $N_4$ whereas the probabilities when the network was misclassified indicated more uncertainty, generally with the probability mass concentrated on both the incorrect and correct networks (with the incorrect network having probabilities in the $50 - 70 \%$ range).

\section{Biological Examples}
\label{subsec:Hybrid}

Quartets and quarnets can play an essential role in the reconstruction of trees and networks since the topological structure of individual quartet trees or networks can encode the topological structure of a larger tree or network from which it is sampled. Given a collection of quartet trees, algorithms such as quartet puzzling, QuartetsMaxcut \cite{snir2012quartet}, and Astral \cite{mirarab2014astral} have different means of selecting the tree which best reflects this information. Network based reconstruction algorithms such as SNaQ \cite{Solis2016} build on the same theoretical framework.

We can summarize such a process for networks with  two key components: estimating the individual quarnet structure and combining the quarnet information into a larger network.  Currently, the QNR-SVM algorithm only addresses the first component of this problem. The algorithm is designed to infer quarnets, but there is no implementation for combining quarnets into a larger network. Unfortunately, most
currently available tools for reconstructing networks do not work with the outputs provided by QNR-SVM.
For example, PhyloNet and SNaQ do not account for the 3-cycle or double-triangle quarnets.
Still, we find that with a reasonable number
of taxa, simple ad hoc approaches
for assembling quarnets can provide a 
useful picture.


\subsection{Description of Primate Data}
In \cite{vanderpool2020primate}, Vanderpool et al.examined the phylogenomic history of 26 primate species along with three non-primate outgroup organisms using various phylogenomic methods designed to reconstruct a species tree.  
In addition, the authors sought to detect interspecific introgression by using a version of the $\Delta$ test \cite{huson2005reconstruction}, an extension of the D-statistic test, more commonly known as the ``ABBA-BABA'' test \cite{durand2011testing, green2010draft, kulathinal2009genomics}. Using these methods, they identified six cases of introgression, primarily among the monkeys in the  \emph{Cercopithecinea} clade. Vanderpool et al. reported that while they detected these introgression events using the $\Delta$ test, they could not get meaningful results using the best-known phylogenomic network methods SNaQ \cite{Solis2016} and PhyloNet \cite{Wen2018}.

Given the need for a formal algorithm for gluing together QNR-SVM estimated quarnets, it is not feasible for us to reconstruct the species tree for all 29 species. Indeed, this would require us to resolve the potential conflicts among all ${29 \choose 4} = 23751$ quarnet networks. So instead, We apply QNR-SVM to two subsets of the primate data representing the \emph{Hominidae}  and \emph{Cercopithecinae} clades.

We used DNA sequence data referenced in the supplementary materials of \cite{vanderpool2020primate} and accessible on Dryad \cite{monkeydata}.
This data contains the concatenated sequence coding sequences of $1730$ single-copy orthologs present in at least $27$ of the $29$ species. This results in $1761114$ bp. 
For each of our selected subsets, we examined only gap-free sites. 
For each subset of four species, we estimated a network using  QNR-SVM and computed $100$ bootstrap networks by resampling among the shared sites. We consider an estimated network to be well supported if it is consistent with at least $95\%$ of the bootstrap networks.  Here we describe each of the examples.\\
\\
\textbf{Example 1: \emph{Hominidae} clade.}
The first data set consists of five primates from the \emph{Hominidae} clade: 
\emph{Pan paniscus}, 
\emph{Pan troglodytes}, 
\emph{Pongo abelii},  
\emph{Gorilla gorilla}, and 
\emph{Homo sapiens} (bonobo, chimpanzee, orangutan, gorilla, and human) and the outgroup \emph{Nomascus leucogenys} (northern white-cheeked gibbon). \\
\\
\textbf{Example 2: \emph{Cercopithecinae} clade.}
The second example consists of the eight primates in the \emph{Cercopithecinae} clade: 
\emph{Chlorocebus sabaeus}, 
\emph{Cercocebus atys}, 
\emph{Mandrillus leucophaeus}, 
\emph{Papio anubis}, 
\emph{Theropithiecus gelada}, 
\emph{Macaca nemestrina}, 
\emph{Macaca fascicularis}, and 
\emph{Macaca mulatta} (green monkey, sooty mangabey, drill, olive baboon, gelada baboon, southern pig-tailed, crab-eating, and rhesus macaques),  as well as the outgroup 
\emph{Colobusangolensis palliatus}
(black and white colobus).

The modern-day macaques are found primarily in southeastern Asia, while the other monkeys are primarily found in Africa. However, their ancestors are thought to have had overlapping territories leading to the potential for introgression \cite{vanderpool2020primate}

\subsection{Results}
\label{subsec:Hybrid results}
The complete listing of QNR-SVM  estimates and associated bootstrap support can be found in {\tt GitHub} repository listed under Supplemental Materials. 

\subsubsection{Example 1: \emph{Hominidae} clade}
The QNR-SVM algorithm returns a tree structure with 100$\%$ bootstrap support for 11 of the 15 subsets of the six samples. An additional tree structure and three network quarnets were estimated, each with boostrap support below 50 out of 100. 

\begin{figure}[h]

\centering
\includegraphics[width = 7cm]{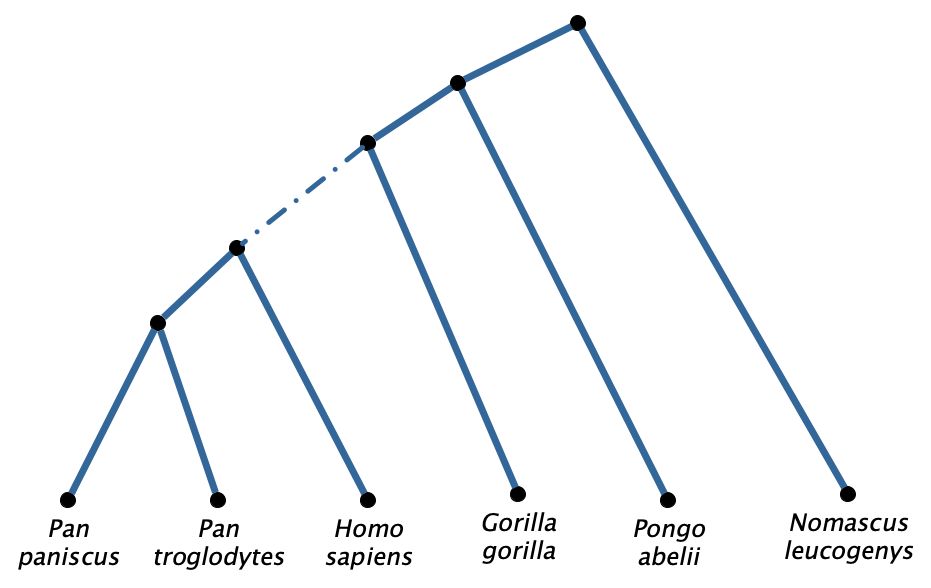}
\caption{Network of the \emph{Hominidae} clade as reconstructed by QNR-SVM. Here the dotted edge indicates possible introgression, though the nature of the suggested event can not be captured with a level-1 network. There is $100\%$ bootstrap support for all of the other edges in the tree.}
\label{fig:apeintro}
\end{figure}

The four estimated quarnets with low bootstrap support all consist of \emph{Homo sapiens}, 
\emph{Gorilla gorilla}, one of the 
\emph{Pan} species as well as either \emph{Pongo} or 
\emph{Nomascus}.  As the \emph{Pan} cherry, 
as well as the (\emph{Pongo}, \emph{Nomascus}) cherry are supported with 100\% bootstrap support across all quarnets, we can combine the four poorly resolved quarnets by examining the relative placement of \emph{Homo}, 
\emph{Gorilla}, \emph{Pan} (ancestral) and (\emph{Pongo}, \emph{Nomascus}) (ancestral). 
\begin{table}
    \centering
    \begin{tabular}{|l|l|}
         (\emph{Pan}, \emph{Homo}),(\emph{Gorilla}, (\emph{Pongo},\emph{Nomascus}) & 32\%  \\
         4-cycle \emph{Homo} as hybrid of (\emph{Pan} and \emph{Pongo}/\emph{Nomascus}) & 36.75\% \\
         4-cycle \emph{Pan} as hybrid of (\emph{Homo} and \emph{Gorilla}) & 21.25\% \\
         4-cycle (\emph{Pongo}/\emph{Nomascus}) as hybrid of (\emph{Pan} and \emph{Gorilla}) & 5.5\% \\
         4-cycle \emph{Gorilla} as hybrid of (\emph{Homo} and \emph{Pongo}/\emph{Nomascus}) & 2.5\% \\
         All Other Topologies & 2 \% \\
 
    \end{tabular}
    \caption{Bootstrap networks estimated by QNR-SVM in placing \emph{Gorilla} and \emph{Homo} between the \emph{Pan} (ancestor) and (\emph{Pongo},\emph{Nomascus}) ancestor.}
    \label{tab:humanchip}
\end{table}
Examining Table~\ref{tab:humanchip}, we see that a total of $66\%$ of the bootstrap networks support a 4-cycle with opposing pairs (\emph{Pan}, \emph{Homo}, \emph{Gorilla}, (\emph{Pongo}/\emph{Nomascus})) with disagreement therein about the direction of introgression. A remaining $32\%$ of the estimated networks support the tree structure (\emph{Pan}, \emph{Homo sapien}),(\emph{Gorilla}, \emph{Pongo}/\emph{Nomascus}).  The relative prevalence of the $4$-cycles indicates a general flow from the (\emph{Gorilla}, \emph{Pongo}, \emph{Nomascus}) group to the (\emph{Pan}, \emph{Pan}, \emph{Homo}) group, which we depict in Figure~\ref{fig:apeintro}.

\subsubsection{Example 2: \emph{Cercopithecinae} clade}
Among the 126 subsets of four samples in the \emph{Cercopithecinae} clade example, the  QMR-SVM algorithm estimated 76 underlying quarnet tree structures and 50 underlying non-tree network structures. Of these 126 estimated quarnets, 71 tree quarnets and nine network quarnets were well supported.

\begin{figure}[h]

\centering
\includegraphics[width = 7cm]{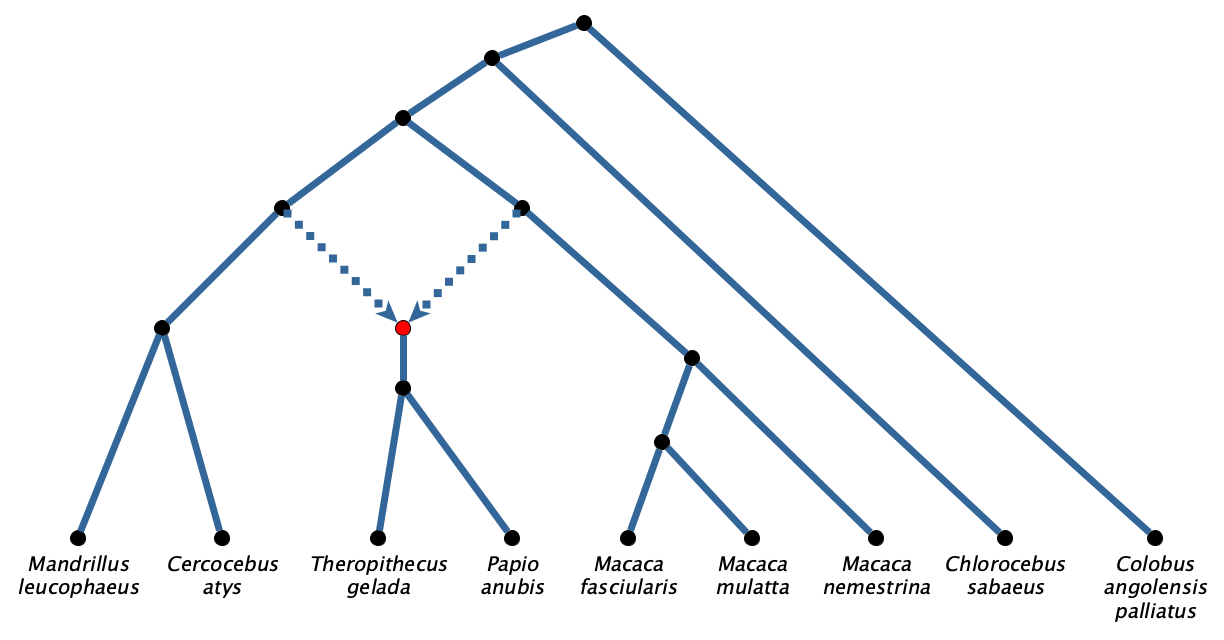}
\caption{Network of the \emph{Cercopithecinae} clade as reconstructed by QNR-SVM.}
\label{fig:monkeycycle}
\end{figure}

While the network structures are not consistent with a single level-one network, the level-one network in Figure~\ref{fig:monkeycycle} is the best estimate. Indeed, $81\%$ of the well-supported quarnet estimates are consistent with the pictured network, and we could not identify a network consistent with a larger proportion of supported quarnets. The majority of the well-supported quarnets that are not compatible with the network in Figure~\ref{fig:monkeycycle} are either 4-cycles whose undirected topologies are compatible with the pictured undirected network or trees that are compatible with a tree-backbone of the pictured network.

In Table~\ref{tab:my_label} we examine the relationship between QNR-SVM estimated a network with when Vanderpool et al. detected introgression using the $\Delta$ test. A quarnet is listed in this table, if either QNR-SVM estimated a well-supported network structure, or Vanderpool et. al found significant evidence of introgression through their application of the $\Delta$ test. In their work, they screened all possible quarnets for evidence of introgression using shorter sequenes for efficiency.  They then tested wether their intial detection of introgression was statistically significant by running the $\Delta$ test on the full sequence. To account for repeated measurements, the authors rejected the null hypothesis that the data could be explained only with a tree when the associated $p$-value was  significant at Dunn-Šidák value $p = 0.00301$. To draw a comparison with their approach, we consider whether QNR-SVM estimated a tree or network, and then indicated statistical support if the associate bootstrap value was at least $95\%$. However, we do not posit a direct equivalence between bootstrap value and confidence level.

{\small
\begin{table}
    \centering
    \begin{tabular}{|c|c|c|c|c|c|}
    \hline
V1	&		V2	&	V3	&	V4	&	QNR-SVM	& $\Delta$	\\
\hline

\emph{M. fascicularis}	&		\emph{M. mulatta}	& \emph{M. nemestrina}	&	\emph{Colobus}	&	tree*	&	 network**	\\
\hline
\emph{M. mulatta}	&		\emph{M. nemestrina}	&	\emph{Theropithecus}	&	\emph{Colobus}	&	tree*	&	network	\\
\hline
\emph{Macaca}	&		\emph{Cercocebus}	&	\emph{Papio}	&	\emph{Colobus}	&	network	&		network	\\
\hline
\emph{Macaca}	&		\emph{Mandrillus}	&	\emph{Theropithecus}	&	\emph{Colobus}	&	network	&	network	\\
\hline
\emph{Macaca}	&		\emph{Cercocebus}	&	\emph{Theropithecus}	&	\emph{Colobus}	&	network	&	network**	\\
\hline
\emph{Macaca}	&		\emph{Mandrillus}	&	\emph{Papio}	&	\emph{Colobus}	&	network	&	network**		\\
\hline
\emph{Cercocebus}	&		\emph{Papio}	&	\emph{Theropithecus}	&	\emph{Colobus}	&	network	& network** \\
\hline

\emph{Macaca}	&		\emph{Mandrillus}	&	\emph{Chlorocebus}	&	\emph{Colobus}	&	network*	&	 tree	\\
\hline
\emph{Mandrillus}	&		\emph{Cercocebus}	&	\emph{Theropithecus}	&	\emph{Colobus}	&	network*	&		tree			\\
\hline
\emph{Mandrillus}	&		\emph{Papio}	&	\emph{Theropithecus}	&	\emph{Chlorocebus}	&	network*	&		tree			\\
\hline

    \end{tabular}
    \caption{Results of introgression testing among samples of four species from the \emph{Cercopithecinae} Clade.  * indicates a QNR-SVM bootstrap support of at least $95\%.$ ** indicates $\Delta$ test significance at  Dunn-Šidák $P = 0.00301$.  Results regarding the $\Delta$ test were taking from S8 Table in the supporting information for \cite{vanderpool2020primate}.}
    \label{tab:my_label}
\end{table}
}

\section{Conclusion and Discussion}
\label{sec:conclusion}

In this manuscript, we explore how algebraic phylogenetic invariants can be paired with support vector machines to infer phylogenetic networks.  As we see in Section \ref{sec:simstudy}, the results on simulated data are promising, as the QNR-SVM algorithm achieves overall accuracy above 85\% on the test set.    Moreover, almost all of the errors fit a predictable pattern in which the predicted network differs from the true network by the insertion or deletion of a reticulation edge in a triangle (or 3-cycle).  While we believe this is a novel and promising approach, it is only a first step in developing an efficient and effective tool for inference.   Indeed, many questions can be explored that are likely to lead to improvements.  We will discuss these questions here, first from the application side, then moving on to the more theoretical side.

In analyzing the \emph{Hominidea} clade, we find non-tree like structure identified by QNR-SVM in the quarnets in Table~\ref{tab:humanchip}.  Inappropriate training data, or a sequencing issue could lead to these findings being an artifact of the QNR-SVM algorithm when the underlying structure is indeed tree-like.  We find this unlikely to be the sole source of the ambiguity, as both more recent and more ancient portions of the phylogeny are clearly captured by QNR-SVM, as are clear relationships among \emph{Homo sapiens} (resp. \emph{Gorilla gorilla}) when analyzed individually with the remaining taxa.  While a more targeted training set (as discussed below) may be helpful in examining this region, we find two alternatives likely more appropriate, both of which highlight the limits of the method.  First it is possible that the underlying network is not level-1, and thus would not be captured in this model.  Even more likely is that the model assumes that data comes from a rooted phylogenetic network which accounts for hybridization, but such a model of hybridization is insufficient to capture introspecific introgression that includes back hybridization. Both of theses scenarios are related to model misspecification, where the first of these highlights the need for better understanding of tree-child networks beyond level-one networks while the second highlights the limits of any approach based on network-based Markov models.

The \emph{Cercopithecinae} clade examination involved the analysis of 126 different quarnets.  The QNR-SVM estimated quarnets support Vanderpool et al.'s findings of significant  introgression among the monkeys in this clade.  Table~\ref{tab:my_label} shows an strong overlap in the collections of monkeys in which we found support for introgression.  Vanderpool et. al represent their introgression findings in a tree diagram overlayed with six  bidirectional arrows connecting edges of the tree in \cite[Figure 4]{vanderpool2020primate}.  In the context of a level-1 network, the same introgression can largely be explained by a single event, as shown in Figure~\ref{fig:monkeycycle}.  More than $80\%$ of the quarnets estimated with at least $95\%$ bootstrap support are consistent with the displayed network, which was constructed using an ad hoc approach to maximizing the quarnet consistency.  Quartet puzzling, quartet MaxCut, and, more recently, the Astral family of algorithms are attempts at efficiently solving the quartet consistency problem in the case of trees.  Widespread implementation of QNR-SVM on large datasets would require similar advances in associated algorithms for networks, with the most challenging components being determining how to resolve differences in conflicting networks and how to estimate the maximum consistency when limited (by time) to examining a subset of the possible quarnet subnetworks. This is perhaps the most significant question from an application point-of-view.

Our next set of questions dive deeper into the methodology of the QNR-SVM algorithm itself --  what modifications can be made to improve the inference of quarnets? The first of these questions regards the set of phylogenetic invariants that we used to transform our data.  As discussed in Section \ref{sec: distinguishing invariants}, one of our main criteria for our method was permutation invariance, and so we constructed a permutation invariant set of phylogenetic invariants.
However, this set contains 1126 polynomials, meaning that our training data is contained in $\mathbb R^{1126}$.  The same method developed with a smaller set of invariants may be as effective and would offer additional savings in the time required to train the model and to classify new observations.
One way to cull the set we used here would be to use some measure of variable importance and retain only those variables with demonstrated power to distinguish between quarnets.  Of course, it would still be desirable for the remaining set of phylogenetic invariants to be permutation invariant.  It may also be possible to use some algebraic or geometric principles to determine theoretically which invariants should perform best at distinguishing between certain quarnets, and then to permute this set to obtain a permutation invariant set of phylogenetic invariants.  Doing so would likely not only improve this method, but could also provide a blueprint for utilizing invariants and algebraic statistics more effectively in model selection.

A second question is how to make the method more robust with respect to the sequence length. 
In our training and test data, each sequence in each alignment consists of $10^6$ sites. 
While we attain some promising results, some decrease in accuracy is expected as we decrease the number of sites since this increases the variance of each invariant residual.
Indeed, when we run the same method where the training and test data consist of sequences of $10^5$ sites, the accuracy drops to 62.38\% (for a test set with 200 test points for each quarnet). This suggests that this method may be more appropriate when there is a sufficient number of sites, for example, when the data consists of a whole genome alignment rather than data from an individual gene  A separate but related question, is how a \emph{mismatch} in the number of sites used in the training and test data affects model accuracy. For example, it may be desirable to pretrain several models on alignments of different lengths (e.g., $10^4$, $10^5$, etc.)  Then, one could choose the model trained on sequences of a similar length to the sequences one wished to classify. We have yet to explore this question in depth, and it may be that models trained on more or fewer sites perform better regardless.

A similar issue to the number of sites is the choice of branch lengths. 
Our experiments suggest the model performs poorly for networks with branch lengths outside of the range of those in the training set. 
One possible fix for this is to retrain the model in the probable range of the branch lengths. 
For example, given a set of aligned sequences, one could use any method to obtain a rough estimate
of the pairwise distances between taxa. Then, the model can be retrained with branch lengths chosen
from intervals that will result in quarnets with the same approximate pairwise distances. 
We noticed better accuracy when we worked on narrower branch length intervals, so in some cases, the computational cost of training the classifier may be worth the improved accuracy  Similar to the above suggestion, one could also pretrain several models on different ranges of branch lengths appropriate for
different sets of taxa. In any case, though, it becomes challenging to infer quarnets with weak
phylogenetic signal as the branch lengths go to zero.

Finally, our method is designed to work with data generated according to the Jukes--Cantor model of DNA sequence evolution on a level-one network. As with many phylogenetic inference methods, it may perform less well when the model is misspecified. In addition, the model only returns level-one networks, so it is not possible for the model to correctly infer the underlying network from data generated by on a network of level-two or greater. 
However, it is likely possible to adapt the method to work with more general group-based models of DNA sequence evolution. For example, several invariants have been found for the level-one quarnets for the Kimura 2-parameter and 3-parameter models \cite{gross2021distinguishing}, so it seems feasible to develop a similar method for these models.  Of course, it may even be possible to train a classifier using a more complicated model with this same set of invariants.  Although this lacks the theoretical justification we offer here, it may be that the invariants here are sufficiently general to distinguish points coming from other Markov models of sequence evolution, for example, the general time-reversible model.



\section*{Supplementary Material}
The supplementary files mentioned in this paper can be found here:\\
\url{https://github.com/lizgross/Inferring-Phylogenetic-Networks-with-QNR-SVM}


\section*{Acknowledgment}

Work on this project was supported by the Mathematical Biosciences Institute and the National Science Foundation under grant DMS-1440386 for CL.  EG was supported by the National Science Foundation under grant DMS-1945584.

\bibliographystyle{plain}
\bibliography{SVM_PhyloInvariants}

\end{document}